\documentclass[%
 reprint,
 amsmath,amssymb,
 aps,
]{revtex4-2}

\usepackage{graphicx}
\usepackage{dcolumn}
\usepackage{bm}
\usepackage{epstopdf} 
\usepackage{hyperref}
\usepackage[mathlines]{lineno}
\usepackage{siunitx}
\usepackage{circuitikz}
\usepackage[dvipsnames]{xcolor}
\begin{document}

\preprint{APS/123-QED}

\title{Low-Gap Hf–HfO$_x$–Hf Josephson Junctions for meV-Scale Particle Detection}

\author{
Y.~Balaji$^{1}$, 
M.~Surendran$^{1}$, 
X.~Li$^{2}$, 
A.~Kemelbay$^{1}$, 
A.~Gashi$^{1}$, 
C.~Salemi$^{2,3}$,
A.~Suzuki$^{2}$, 
S.~Aloni$^{1}$,
A.~Tynes~Hammack$^{1}$, 
and A.~Schwartzberg$^{1}$\thanks{amschwartzberg@lbl.gov}
}

\affiliation{$^{1}$Molecular Foundry, Lawrence Berkeley National Laboratory, Berkeley, CA-94720, USA}
\affiliation{$^{2}$Physics Division, Lawrence Berkeley National Laboratory, Berkeley, CA-94720, USA}
\affiliation{$^{3}$Department of Physics, University of California Berkeley, Berkeley, CA-94720, USA}

\date{\today}

\begin{abstract}
Superconducting qubits have motivated the exploration of Josephson-junction technologies beyond quantum computing, with emerging applications in low-energy photon and phonon detection for astrophysics and dark matter searches. Achieving sensitivity at the THz (meV) scale requires materials with smaller superconducting gaps than those of conventional aluminum or niobium-based devices. Here, we report the fabrication and characterization of hafnium (Hf)–based Josephson junctions (Hf–HfO$_x$–Hf), demonstrating Hf as a promising low-$T_c$ material platform for ultra-low threshold single THz photon and single-phonon detection. Structural and chemical analyses reveal crystalline films and well-defined oxide barriers, while electrical transport measurements at both room and cryogenic temperatures exhibit clear Josephson behavior, enabling extraction of key junction parameters such as critical current, superconducting gap and normal-state resistance. This work presents the first comprehensive study of Hf-based junctions and their potential for next-generation superconducting detectors and qubit architectures leveraging low superconducting gap energies.

\end{abstract}

\maketitle

Progress in superconducting qubits has been driven by advances in device fabrication, materials engineering, and noise mitigation~\cite{Oliver2013, Gambetta2017, Place2021}. 
Beyond quantum computing, these qubits are now being leveraged for applications in minuscule signal detection, such as single THz photon detection~\cite{echternach2018single} for astrophysics observation and axion like particle or dark photon searches, and athermal phonon detection for light dark matter searches~\cite{Fink2024,Ramanathan2024}. Conventional detectors such as microwave kinetic inductance detectors (MKIDs) and transition-edge sensors (TESs) exhibit excellent sensitivity, but their energy thresholds typically remain at the eV scale~\cite{Day2003, IrwinHilton2005}. 
To extend the sensitivity of single-photon counting axion searches to frequencies as low as $O(100)\si{\giga\hertz}$~\cite{chiles2022new, liu2022broadband} and the sensitivity of phonon-mediated detectors to dark matter masses as low as $O(10)\si{\kilo\electronvolt}$~\cite{knapen2018detection, griffin2020multichannel}, \si{\milli\electronvolt}-scale sensitivity is required. Josephson-junction (JJ)–based quantum sensors have emerged as a promising alternative~\cite{Kurinsky2020}. 
Recently developed qubit-derived detectors—such as the Quantum Capacitance Detector (QCD)~\cite{Shaw2009}, Quantum Parity Detector (QPD)~\cite{Ramanathan2024}, and Superconducting Quasiparticle-Amplifying Transmon (SQUAT)~\cite{Fink2024}— leverage the intrinsic single-quasiparticle sensitivity of superconducting junctions. 
In these devices, absorbed energy generates non-equilibrium quasiparticles that tunnel across the JJ, producing measurable changes in the qubit’s parity or resonance frequency. 

Using superconductors with lower critical temperature, $T_c$ and smaller superconducting gap enhances detector responsivity by increasing the quasiparticle population and lifetime per absorbed energy~\cite{Catelani2011}.
More importantly, building an effective phonon-sensing qubit requires additional large phonon absorbers, typically $O(100)\si{\mu\meter}$ long aluminum (Al) pads in contact with the JJ, which allows the excited quasiparticles to diffuse to the vicinity of the JJ. Al is preferred for its extremely long quasiparticle lifetime~\cite{kaplan1976quasiparticle} and good acoustic impedance matching to common substrates. 
When the JJ uses a superconductor with a $T_c$ lower than that of Al (Al $T_c$ = 1.2~K), it naturally forms a quasiparticle trap, concentrating excited quasiparticles within the small junction volume. The trap significantly increases the signal quasiparticle density and the tunneling probability. It also resolves the conflicting requirements of high phonon collection efficiency favored by large absorber volume and high responsivity with smaller superconductor volume~\cite{Fink2024}. The quasiparticle-trapping technique has been successfully demonstrated in tungsten TES with Al absorbers~\cite{fink2020characterizing, chang2025first}. For traps formed entirely of superconducting metals on both sides, the $T_c$ ratio needs to be greater than $4$ to insure effective trapping~\cite{kaplan1976quasiparticle}. Thus, simply reversing the configuration of a gap-engineered Al junction~\cite{mcewen2024resisting} is insufficient, as gap engineering would require modifying the Al thickness to be below 30~nm, unfavorable for  phonon collection nor quasiparticle diffusion.

To date, JJ technologies have primarily relied on Al and niobium, while junctions fabricated from lower-$T_c$ superconductors remain largely unexplored.
Hafnium (Hf) is a promising candidate for a low-$T_c$ superconductor. Its bulk $T_c$ is near 128~mK~\cite{kraft1998hafnium},with a London penetration depth of about 20 nm~\cite{kraft1998hafnium}. For a 125 nm-thick film, the surface kinetic inductance is relatively large, on the order of 15--20~pH/$\square$~\cite{Rotermund2024}, making Hf particularly suitable for enhancing photon and phonon energy responsivity through increased quasiparticle excitation per absorbed energy, as well as for effective quasiparticle trapping in phonon sensing applications. Although Hf has been investigated as a material for superconducting tunnel junctions (STJs), these studies have not yet led to the successful realization of functional Hf-based JJs.~\cite{kraft1998hafnium,Kim2012}
\begin{figure*}
    \centering
    \includegraphics[width=0.7\textwidth]{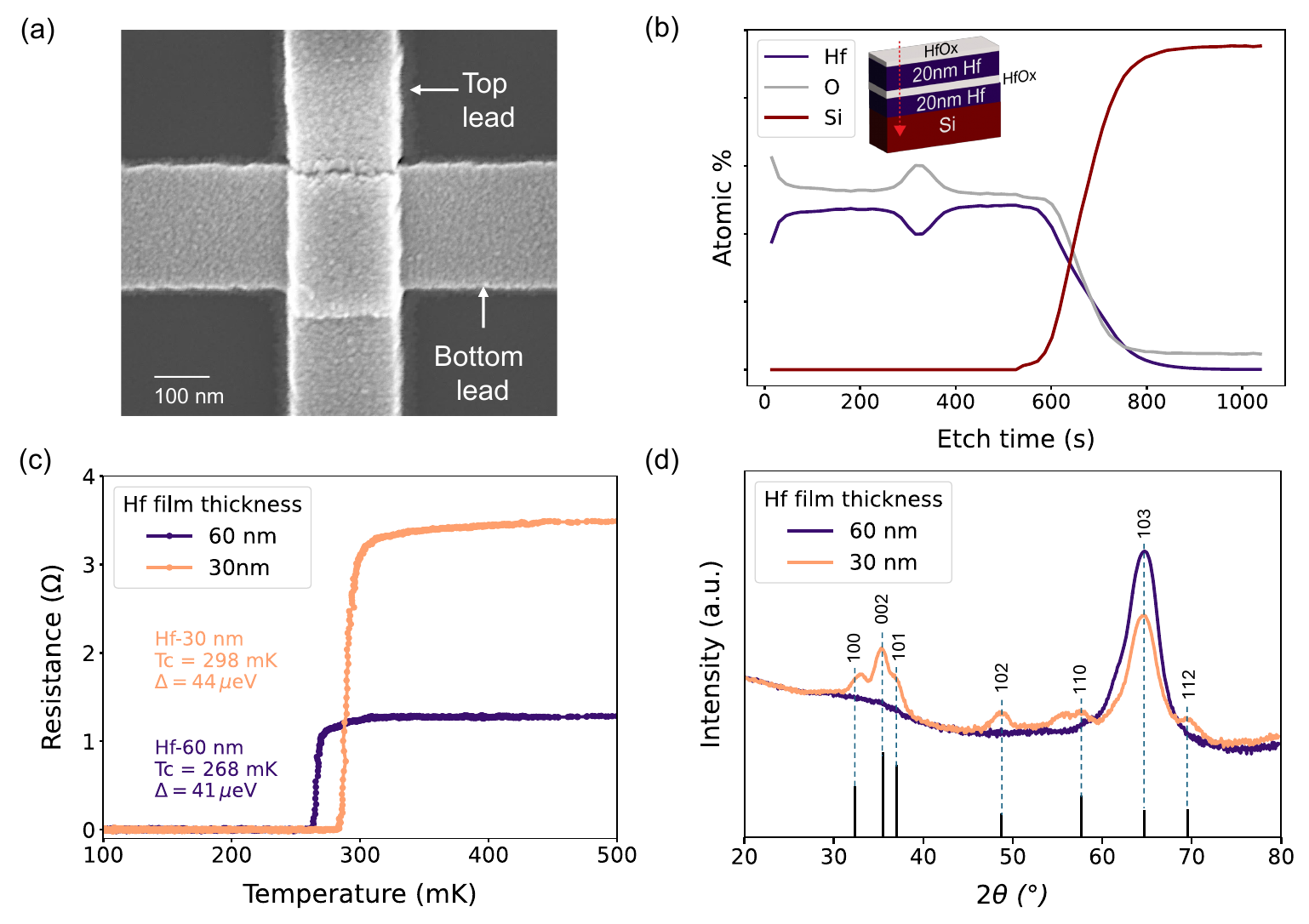}
    \caption{\label{fig:1}(a) Top-view SEM image of the JJ overlap region. (b) XPS depth profile of a blanket Hf-HfO$_x$-Hf stack on a Si substrate, showing the evolution of relative atomic percentage with increasing etch time. The etch time corresponds to the composition from top to bottom as shown in schematic of the thin film stack.(c) Resistance measurements as a function of temperature to extract the $T_c$ of  30 nm and 60 nm Hf films. (d) Grazing incidence XRD patterns of 30~nm and 60~nm Hf films on Si.}
\end{figure*}

In this work, we demonstrate the fabrication of Hf–HfO$_x$–Hf JJs, establishing Hf as a viable platform low-$T_c$ superconducting devices for meV-scale detection. 

We present the first comprehensive materials and electrical characterization of Hf JJs, highlighting their potential for next-generation low-energy quantum sensors. The junctions were characterized using scanning electron microscopy (SEM) and atomic force microscopy (AFM) to determine geometry and layer thicknesses. The crystal structure and chemical composition of Hf films were characterized using X-ray diffraction (XRD) and X-ray photoelectron spectroscopy (XPS) respectively. Electrical performance was evaluated through current–voltage ($I–V$) measurements conducted at both room and millikelvin temperatures to extract the critical current ($I_c$), normal resistance ($R_N$) and $T_c$. These results identify Hf as a promising material platform for next-generation Josephson devices, supporting enhanced sensitivity in superconducting detectors.

The JJs were fabricated using the well-established double-angle shadow evaporation technique commonly employed in Al-based qubits~\cite{Muthusubramanian2024}. Test junctions were patterned by e-beam lithography to define the JJ leads and contact pads, followed by e-beam evaporation and an oxidation step to form the Hf leads and the HfO$_x$ tunnel barrier~(see Supplementary sec~\ref{Sup:devicefab} for fabrication details).   Fig.~\ref{fig:1}(a) shows a SEM image of a fabricated JJ, with lead widths of approximately 200~nm. The JJs employ a Manhattan-style  geometry with varying  junction overlap areas~\cite{Kreikebaum2020}. The Hf grain size was measured to be below 10~nm. To obtain the height profile, AFM measurements were performed and the bottom and top lead thicknesses were measured to be 30~nm and 60~nm respectively. (Supplementary sec~\ref{Sup:SemAfm} for SEM and AFM details)

To confirm the presence of the HfO$_x$ tunnel barrier layer in the JJ, we fabricated a representative trilayer stack consisting of a thin HfO$_x$ layer sandwiched between two 20 nm Hf films. XPS depth profiling was performed on this stack to estimate the HfO$_x$ thickness and to identify the oxidation state of Hf within the oxide. 
Fig.~\ref{fig:1}(b) presents the compositional depth profile of Hf$\,{4f}$, O$\,{1s}$, Si$\,{2p}$ peaks as a function of etch time with a schematic of Hf-HfO$_x$-Hf cross-section shown in the inset. A distinct increase in oxygen concentration, accompanied by a corresponding decrease in hafnium concentration, is observed within the oxide region. From the XPS depth profiling, the thickness of the HfO$_x$ layer is estimated to be approximately 4–5~nm. A significant oxygen concentration is also observed within the bulk of the metal, which may arise from interstitial non-bonded oxygen incorporated during film growth, as well as residual background oxygen from the XPS analysis chamber. This is consistent with the absence of HfO$_x$-related peaks in the Hf$,{4f}$ spectra away from the sandwiched HfO$_x$ layer. The evolution of the Hf$,{4f}$ spectra in Supplementary Sec.~\ref{Sup:XPS} also highlights the presence of HfO$_x$ layer, with its characteristic peaks appearing and disappearing as etching progresses.

To investigate the structural and electrical properties of Hf layers in the JJ, we deposited blanket Hf films with thicknesses corresponding to those of the bottom and top junction electrodes (30~nm and 60~nm respectively). $T_c$ measurements were  performed by measuring the resistance as a function of temperature. As shown in Fig.~\ref{fig:1}(c), the measured transition temperatures are $T_{c,30} = 298$~mK for the 30~nm film and $T_{c,60} = 268$~mK for the 60~nm film. The superconducting energy gap ($\Delta$) was extracted using the BCS relation  $\Delta = 1.76\,k_B T_c$~\cite{Bardeen1957}, yielding $\Delta_{30nm} = 44~\mu$eV and $\Delta_{60nm} = 41~\mu$eV~(see Supplementary sec~\ref{Sup:Tc} for $T_c$ measurement details)

Grazing-incidence XRD measurements of the Hf films confirmed that they are polycrystalline with a hexagonal close-packed (hcp) structure. Fig.~\ref{fig:1}(d) shows XRD scans for the 30 nm and 60 nm Hf films, along with the reference hcp Hf pattern ($P6_3/mmc$). The 30 nm film exhibits multiple distinct reflections corresponding to the hcp Hf phase, whereas the 60 nm film appears textured, with only the 103 reflection detected. The broad diffraction peaks indicate a small grain size. Notably, no HfO$_2$-related peaks are observed, suggesting minimal oxidation in the as-deposited films.

\begin{figure}[t]
    \centering
    \includegraphics[width=0.35\textwidth]{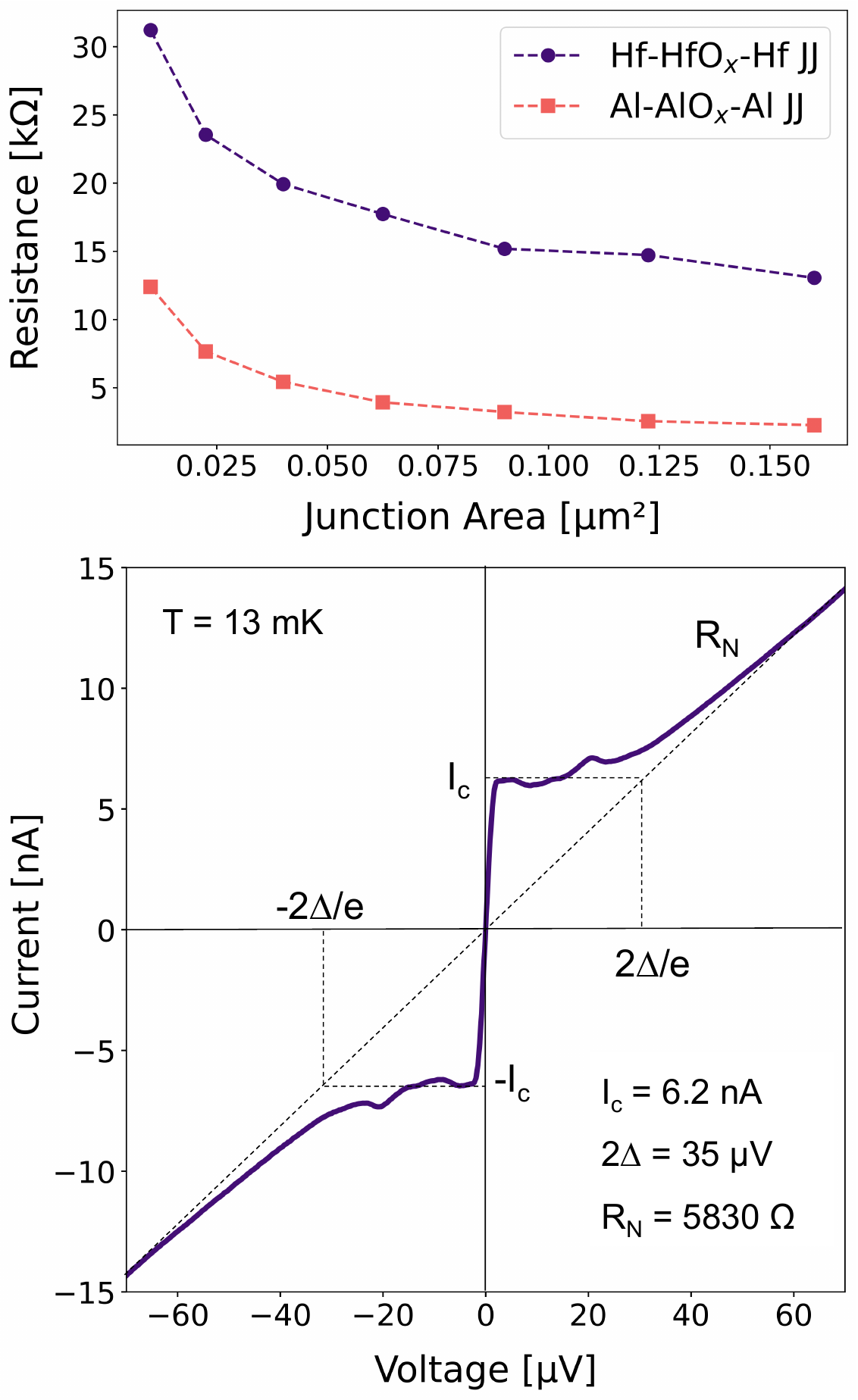}
    \caption{\label{fig:2}(a) Room temperature resistance measurements with varying dimensions. The plot shows the resistance of Hf JJ along with that of a standard Al JJ, with the Hf JJ resistance up to 2-5 times higher than Al JJ (b) $I–V$ characteristics of the Hf JJ measured at 13~mK. The plot shows the extracted $I_c$, $\Delta$ and $R_N$.}
\end{figure}

The electrical performance of the JJs was first evaluated at room temperature. Fig.~\ref{fig:2}(a) shows the room-temperature resistance of JJs with varying junction overlap areas. For comparison, the resistance of an Al–AlO$_x$–Al JJ fabricated using the same fabrication steps and under identical oxidation conditions is also shown. The Hf JJs exhibit resistances that are approximately 2–5 times higher than those of the Al JJs. This difference may arise from several factors, including variations in film resistivity and differences in the tunnel barrier thickness. 

To confirm the fabricated devices operate as a JJ, we extracted the superconducting parameters such as  $I_c$ and $\Delta$. $I$–$V$ measurements were carried out in a dilution refrigerator at a base temperature of $13~\mathrm{mK}$. The junction was connected in series with a voltage divider circuit to precisely control the voltage sweep. The resulting current was measured using a DC SQUID amplifier as the readout (supplementary sec~\ref{Sup:DFsetup} for measurement details). Fig.~\ref{fig:2}(b) shows the $I$–$V$ characteristics of a JJ with an overlap area of $0.16~\mu\mathrm{m}^2$, with dashed lines highlighting the different parts of the curve. The extracted critical current is $I_c = 6.2~\mathrm{nA}$. Beyond a certain voltage, the transport in the junction is dominated by quasiparticle tunneling, resulting in a linear behavior corresponding to the normal-state resistance, $R_N = 5.8~\mathrm{k\Omega}$. 
The intermediate region between $I_c$ and $R_N$ corresponds to the subgap regime, which extends up to approximately twice the superconducting gap, $2\Delta \approx 35~\mu\text{eV}$. In this region, the presence of in-gap states leads to a finite density of states at energies below the superconducting gap. Based on the Ambegaokar–Baratoff (AB) relation~\cite{Ambegaokar1963}, the expected critical current for a tunnel junction is given by
\begin{equation}
I_c R_N = \frac{\pi \Delta}{2e} \tanh\left(\frac{\Delta}{2 k_{\mathrm{B}} T}\right),
\label{eq:AB}
\end{equation}

\begin{figure*}
    \centering
    \includegraphics[width=0.65\textwidth]{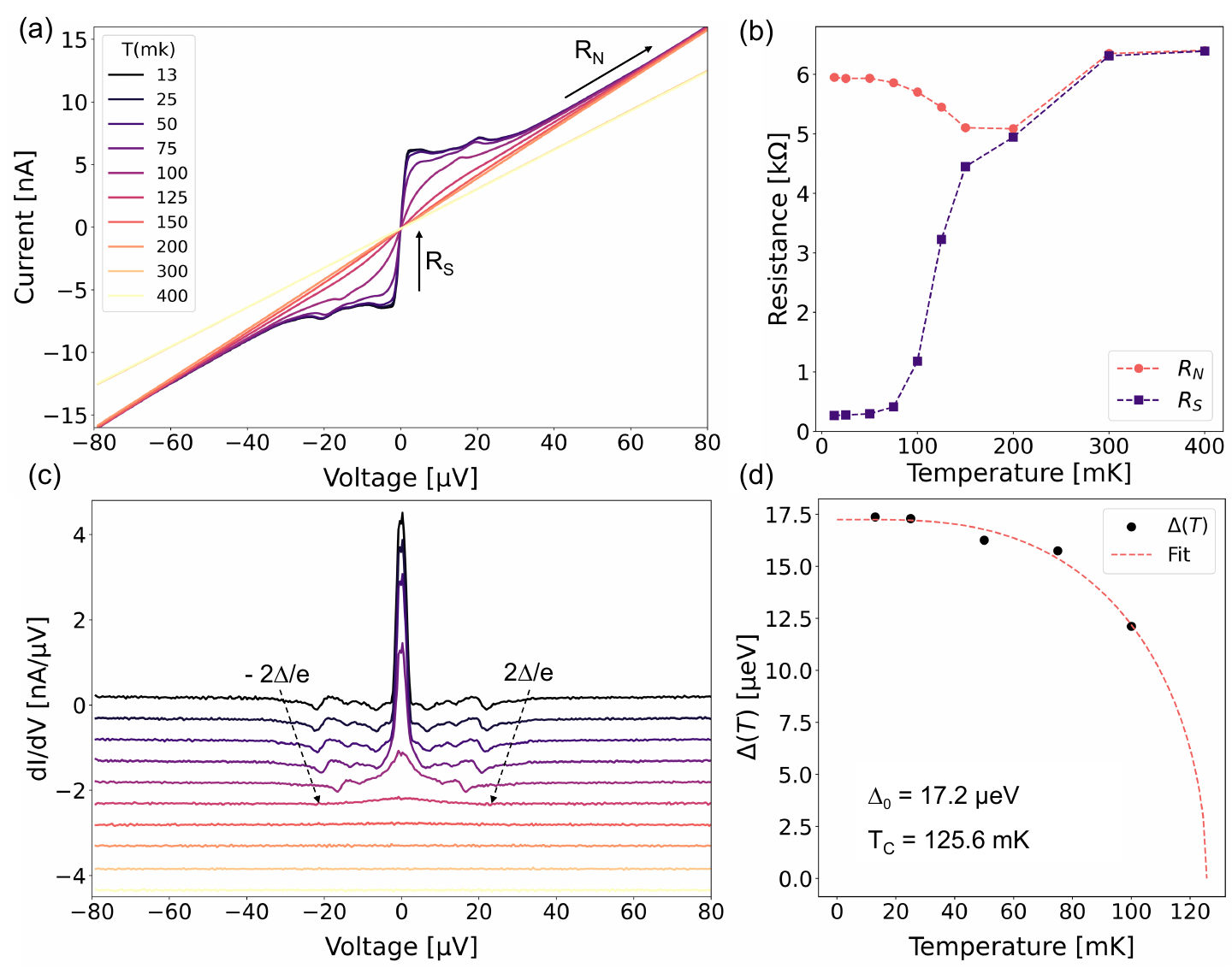}
    \caption{\label{fig:3}(a) Temperature dependent $I-V$ characteristic of the JJ showing that the supercurrent reduces with increasing temperature. (b) The supercurrent resistance ($R_S$) and normal resistance ($R_N$) as a function of temperature. The resistance values of R$_S$ and R$_N$ are extracted from the slope of $I-V$ curves marked in (a). (c) Numerical differentiation of the temperature dependent $I-V$ curve. The dashed line corresponds to measured gap (2$\Delta$) region. (d) The extracted gap from (c) as a function of temperature. The fit is used to extract the gap at 0K ($\Delta_0$) and the JJ's critical temperature T$_c$.}
\end{figure*}

where $\Delta$ is the superconducting gap, $e$ is the elementary charge, and $T$ is the electronic temperature, representing the temperature of the quasiparticle distribution in the superconducting electrodes. Although it is assumed that the electronic temperature is well thermalized with the bath temperature ($T = 13~\mathrm{mK}$), we observe that the electron temperature decouples from the cryostat temperature below approximately 25 mK, where the $I$–$V$ characteristics does not show noticeable temperature dependence. Further investigation of this decoupling is required in the future to accurately determine the effective JJ temperatures. By substituting the measured values of $I_c$ and $R_N$ into the AB relation, we obtain $\Delta_{\mathrm{AB}} = 23~\mu\text{eV}$. However, the experimentally extracted gap from the $I–V$ measurement is $\Delta_{\mathrm{meas}} \approx 17.5~\mu\text{eV}$. This discrepancy likely arises from subgap states in the tunnel barrier, induced by disorder or defects at the Hf–HfO$_x$ interfaces thereby reducing the overall superconducting gap of the JJ. These subgap states can be described by multiple Andreev reflections (MAR)~\cite{Bardeen1957} and by Dynes broadening, which accounts for finite quasiparticle lifetimes~\cite{Dynes1978}. The Dynes parameter $\gamma$ can be estimated from the ratio of the normal-state resistance $R_N$ to the subgap resistance $R_{\mathrm{subgap}}$ as
\begin{equation}
\gamma = \Delta \times \frac{R_N}{R_{\mathrm{subgap}}},
\end{equation}
yielding $\gamma = 7.2~\mu\text{eV}$ for this device. The presence of these states leads to a smearing of the coherence peaks and a finite density of states within the superconducting gap.

Fig.~\ref{fig:3}(a) shows the $I$–$V$ characteristics of the JJ measured at temperatures ranging from $13~\mathrm{mK}$ to $400~\mathrm{mK}$.  Fig.~\ref{fig:3}(b) plots the resistances extracted at the supercurrent, $R_S$ and normal current, $R_N$ regions of the curves. $R_S$ increases with temperature due to increased quasiparticle conduction in the super current regime. We observe that $R_N$ decreases as the temperature increases from 13~mK to approximately 200~mK, before increasing again at higher temperatures. This behavior can be attributed to the junction being in a mixed state, where part of the current is carried by Cooper pairs, providing a low-resistance channel, while thermally excited quasiparticles contribute a resistive component. As the temperature approaches $T_c$, the supercurrent weakens, but thermal activation facilitates Cooper pair tunneling, slightly reducing the measured resistance. This reduction in resistance corresponds directly to an increase in the slope of the supercurrent branch, reflecting the gradual onset of dissipative processes. Above $T_c$, superconductivity is suppressed, and the junction behaves as a normal metal, with resistance increasing due to phonon scattering. 

Fig.~\ref{fig:3}(c) shows the numerical differential, d$I$/d$V$, of the data presented in Fig.~\ref{fig:3}(a), enabling a more accurate determination of $\Delta$, as indicated by the dashed black line. We observe that the sub-gap states and $\Delta$ decrease with increasing temperature due to thermally activated quasi particle tunneling.  The extracted gap values were plotted as a function of temperature in Fig.~\ref{fig:3}(c) and fitted using the standard BCS relation between the superconducting gap and critical temperature:

\begin{equation}
    \Delta(T) = \Delta_0 \tanh\!\left[1.74 \sqrt{\frac{T_c}{T} - 1}\,\right],
\end{equation}

where $\Delta_0$ is the superconducting gap at zero temperature and $T_c$ is the critical temperature. From this fit, we obtain $\Delta_0 = 17.24~\mu\mathrm{eV}$ and $T_c = 125~\mathrm{mK}$ for our Hf JJ. We notice that the Tc of the JJ is lower than the Tc of the individual Hf films. This is probably due to the presence of  sub gap states present in the tunnel barrier as well as contribution to other lossy interfaces arising during the JJ fabrication~\cite{Im2010}. Table~\ref{tab:tab_1} shows the different parameters extracted from the junction measurements.

\begin{table}[h!]
\centering
\caption{\label{tab:tab_1} Parameters of the measured Hf JJ.}
\begin{tabular}{l c c}
\hline
\textbf{Parameter} & \textbf{Symbol} & \textbf{Value} \\
\hline
Critical current & $I_c$ & 6.2~nA \\
Superconducting gap & $\Delta$ & 17.2~$\mu$eV \\
Normal-state resistance & $R_N$ & 5.2~k$\Omega$ \\
Critical temperature & $T_c$ & 0.125~K \\
Operating temperature & $T_{\mathrm{op}}$ & 0.013~K \\
Josephson energy & $E_J/h$ & 3~GHz \\
Dynes parameter & $\gamma$ ($\lambda$) & 7.2~$\mu$eV \\
Junction area & $A_{\mathrm{JJ}}$ & 0.16~$\mu$m$^2$ \\
\hline
\end{tabular}
\label{tab:JJparams}
\end{table}

In summary, we have demonstrated the fabrication and characterization of hafnium-based Josephson junctions. Material characterization confirms that HfO$_x$ forms a reliable and stable tunnel barrier capable of supporting the Josephson effect.  The measured critical current and energy gap values align well with the requirements of various qubit architectures and detection applications. This novel device demonstrates the potential of low-gap superconductors as a promising platform for achieving meV-scale quasiparticle detection. Our ongoing work focuses on further characterizing the oxide tunnel barrier by varying the oxide growth dynamics and investigating the influence of junction area on the device’s critical current.

\begin{acknowledgments}
This work was supported by the Department of Energy, Office of Science, Office of Basic Energy Sciences, Materials Sciences and Engineering Division under Contract No. DE-AC02-05-CH11231 in the Phonon Control for Next Generation Superconducting Systems and Sensors FWP (KCAS23).
Work at the Physics Division was supported by the Office of Science, Office of High Energy Physics, of the U.S. Department of Energy under Contract- DE-AC02-05CH11231 (KA2501032).
 We would like to thank Scott Dhuey for performing the e-beam patterning during the sample lithography process. We would like to thank Kungang Li, Kaja Rotermund and Rebecca Carney for assisting with the cryogenic measurements.

\end{acknowledgments}



%


\clearpage
\onecolumngrid   

\begin{center}
    \textbf{\large Supplementary Information for ``Low-Gap Hf–HfO$_x$–Hf Josephson Junctions for meV-Scale Particle Detection''}
\end{center}

\bigskip

\noindent
Y.~Balaji$^{1}$, 
M.~Surendran$^{1}$, 
X.~Li$^{3}$, 
A.~Kemelbay$^{1}$, 
A.~Gashi$^{1}$, 
C.~Salemi$^{2,3}$,
A.~Suzuki$^{2}$, 
S.~Aloni$^{1}$,
A.~Tynes~Hammack,$^{1}$, 
and A.~Schwartzberg$^{1}$\\
\textit{$^{1}$Molecular Foundry, Lawrence Berkeley National Laboratory, 1 Cyclotron Road, Berkeley, CA 94720, USA}\\
\textit{$^{2}$Physics Division, Lawrence Berkeley National Laboratory, 1 Cyclotron Road, Berkeley, CA 94720, USA}\\
\textit{$^{3}$Department of Physics, University of California Berkeley, Berkeley, CA-94720, USA}\\

\setcounter{section}{0}
\setcounter{subsection}{0}
\setcounter{figure}{0}
\setcounter{table}{0}
\setcounter{equation}{0}
\renewcommand{\thefigure}{S\arabic{figure}}
\renewcommand{\thetable}{S\arabic{table}}

\section{Device Fabrication}\label{Sup:devicefab}

For junction fabrication, the silicon wafer was first spun with MMA e-beam resist at a speed of \(4000 \, \text{RPM}\) for \(45 \, \text{s}\) and hard baked for \(3 \, \text{min}\) at \(180^{\circ}\text{C}\). Subsequently, PMMA e-beam resist was spun at \(2000 \, \text{RPM}\) for \(45 \, \text{s}\) and hard baked for \(3 \, \text{min}\) at \(180^{\circ}\text{C}\). The resist was then patterned using electron-beam lithography (Raith EBPG). Development was carried out in a MIBK/IPA (1:3) solution at room temperature for \(60 \, \text{s}\), followed by immersion in IPA for \(30 \, \text{s}\) and N\(_2\) blow drying.

Following lithography, the wafer was mounted into the e-beam evaporation system (Angstrom Engineering), which is integrated within a cluster tool system. For junction deposition, \(30 \, \text{nm}\) of Hf was evaporated at a \(45^{\circ}\) angle, followed by an oxidation step. The second Hf lead was subsequently deposited at a \(45^{\circ}\) angle with a thickness of \(70 \, \text{nm}\). The pressure in the evaporation chamber during deposition was around \(1 \times 10^{-7} \, \text{torr}\).

Liftoff was performed using N-Methyl-2-pyrrolidone (NMP) for \(3 \, \text{h}\) at \(80^{\circ}\text{C}\) on a hot plate, followed by cleaning in IPA and \(N_2\) blow drying. 
The junction pads were wire-bonded to the PCB setup to establish electrical contacts.

\section{Material Characterization}

\subsection{SEM and AFM characterization}\label{Sup:SemAfm}
\begin{figure*}[h!]
    \centering
    \includegraphics[width=0.8\textwidth]{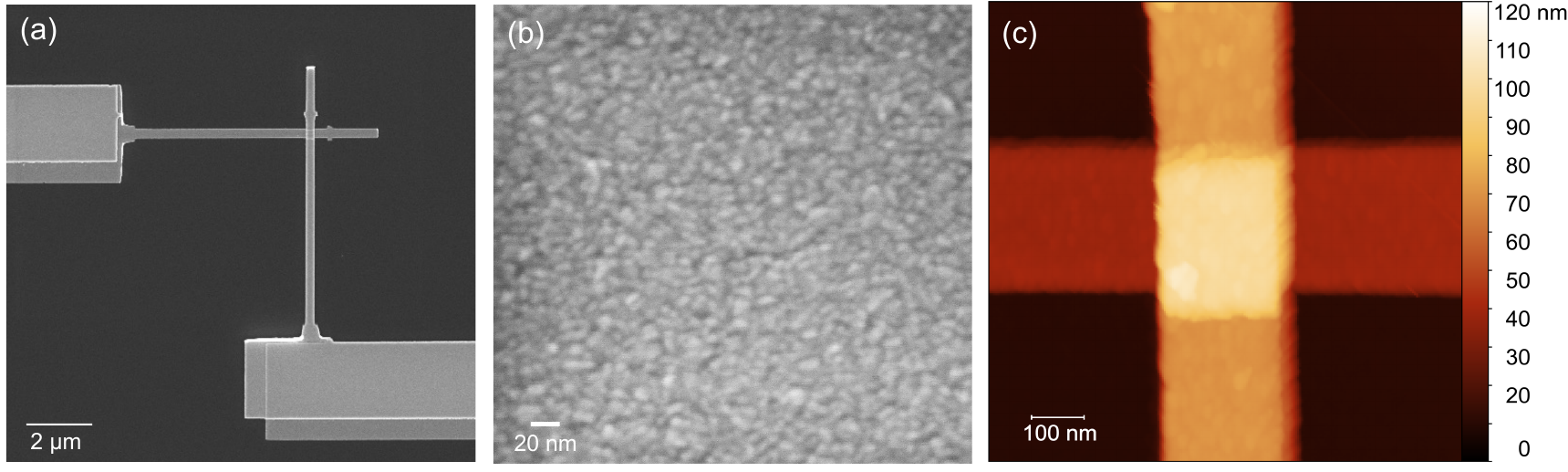}
    \caption{\label{figS1}(a) SEM image of the Hf JJ fabricated using the Manhattan style geometry (b) SEM image showing the grain size of the Hf film. The grains sizes are below 10~nm. (c) AFM image of the Hf JJ.}
\end{figure*}

SEM was performed using a Zeiss Ultra 60 field emission scanning electron microscope at 5kV using an in-lens detector. Fig.~\ref{figS1}(a) shows the SEM image of 200~nm wide fabricated JJ test structures. Fig.~\ref{figS1}(b) shows the Hf grains with grain sizes of less than 10~nm. 

To measure surface roughness and height of the junction electrodes, AFM was performed using a  Bruker Dimension ICON AFM, using the ScanAsyst peak force tapping mode. Fig.~\ref{figS1}(c) show the AFM image of 200~nm wide JJ. The bottom and top leads were measured to be 30~nm and 60~nm respectively.

\subsection{XPS characterization}\label{Sup:XPS}
X-ray photoelectron spectroscopy (XPS) was performed using a Thermo-Fisher K-Alpha+ XPS equipped with a focused monochromatic Al X-ray source, a collection lens with a \(30^\circ\) half-angle acceptance, and a hemispherical analyzer with a multichannel detector at normal incidence to the sample. The x-ray spot size used was 400 $\mu$m. A dual-mode monatomic argon ion source was used for sputter cleaning and depth profiling. A Shirley inelastic scattering background was used for Hf${4f}$, O${1s}$, Si${2p}$ peaks. A total of 80 points were collected for the dept profiling.

\begin{figure*}[h!]
    \centering
    \includegraphics[width=0.3\textwidth]{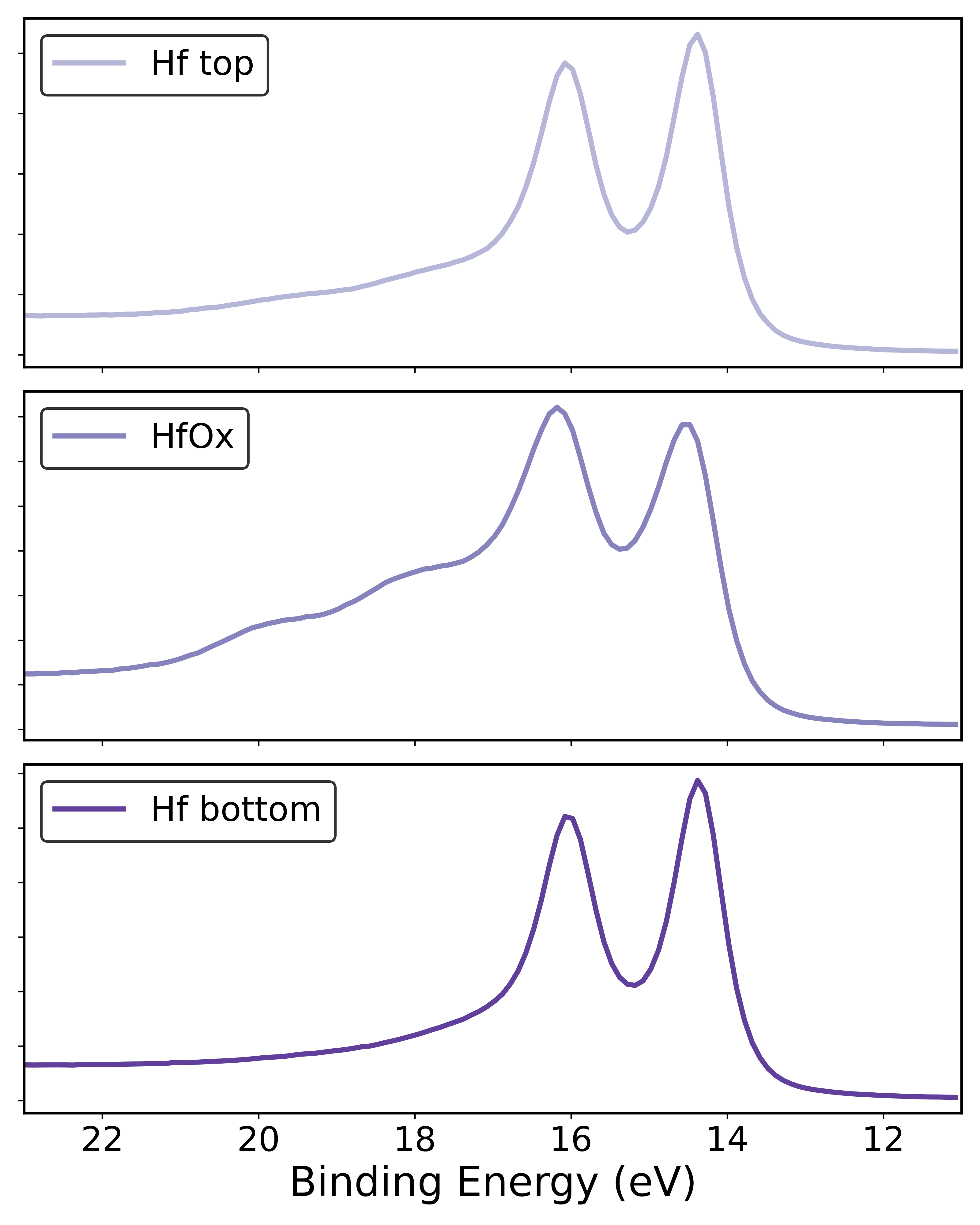}
    \caption{\label{figS2}Hf$\,\mathrm{4f}$ XPS spectra from a blanket Hf-HfO$_x$-Hf stack on a Si substrate at various etch times. }
\end{figure*}

Fig.~\ref{figS2} shows the Hf$\,\mathrm{4f}$ spectra corresponding to the top Hf layer (etch time = 105 s), the intermediate HfO$_x$ layer (etch time = 315 s), and the bottom Hf layer (etch time = 525 s). The appearance of a clear Hf$^{4+}$ component characteristic of stoichiometric HfO$_2$ indicates the formation of a chemically uniform oxide, which is essential for achieving a non-leaky junction with a uniform barrier height.

\subsection{XRD of Hafnium films}\label{Sup:XRD}
The grazing incidence XRD measurements of the Hf films were conducted using a Rigaku SmartLab X-ray Diffractometer with a Cu K$\alpha$  radiation ($\lambda$ =$1.5418~\mathrm{\AA}$) source in a grazing incident angle geometry with a 1D detector.

\section{Measurement Setup}\label{Sup:DFsetup}

\subsection{Room temperature resistance measurement}

Room temperature measurements were performed using the SEMIPROBE PS4L semi-automated probe station. $I-V$ sweeps were performed on JJ test structures using a Keithley 2450 sourcemeter and the resistance is extracted from the slope of the $I-V$ curve.

\subsection{Cryogenic measurements}
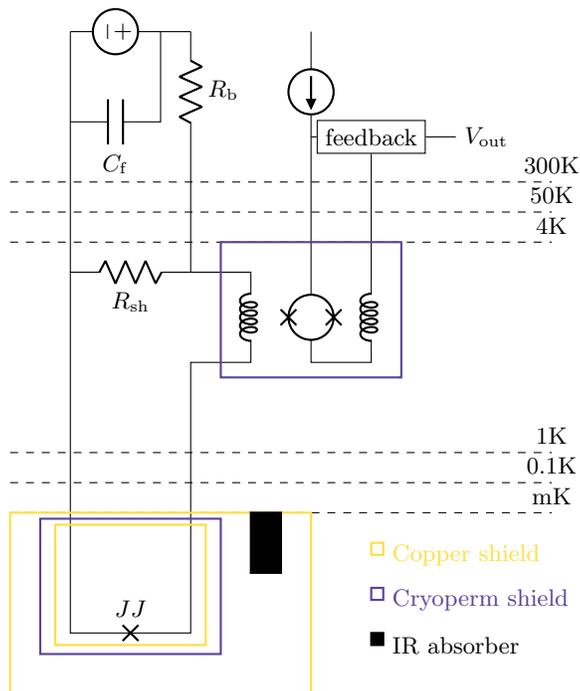
\begin{figure}
    \centering
    \ctikzset{bipoles/length = 1cm}
\begin{tikzpicture}
\begin{scope}[scale=0.8]
    \draw (0,4) --++(-0.5,0) to[american voltage source]++(-1.5,0);
    \draw (-0.5,4) --++(0,-1.5) to[C=$C_\mathrm{f}$] ++(-1.5,0);
    \draw (0,4) to [R=$R_\mathrm{b}$]++(0,-2) -- ++(0,-2) to [R=$R_\mathrm{sh}$] ++ (-2,0) --++(0,4);
    \draw (-2,0) --++(0,-6) to [barrier=$JJ$] ++ (2,0) --++(0,4.5) --++(1,0) to [L] ++ (0,1.5) --++(-1,0);
    \draw (3,2.25) node[draw] (feedback) {feedback}
    (2,4) to[american current source]++(0,-2) -- (2,0) to [squid] ++ (0,-1.5) -- (3,-1.5) to [L] ++(0,1.5)  -- (feedback.south);
    \draw (2,2.25) -- (feedback.west);
    \draw (2.5,-3) ;
    \draw (feedback.east) --++(0.5,0) node[anchor=west]{$V_\mathrm{out}$};
    \draw[dashed] (-3,1.5) --++(9,0) node[anchor=south]{300K};
    \draw[dashed] (-3,1) --++(9,0) node[anchor=south]{50K};
    \draw[dashed] (-3,0.5) --++(9,0) node[anchor=south]{4K};
    \draw[dashed] (-3,-3) --++(9,0) node[anchor=south]{1K};
    \draw[dashed] (-3,-3.5) --++(9,0) node[anchor=south]{0.1K};
    \draw[dashed] (-3,-4) --++(9,0) node[anchor=south]{mK};
    \draw[Violet,thick] (-2.5,-4.1) rectangle ++(3,-2.25);
    \draw[Violet,thick] (0.5,0.5) rectangle ++(3,-2.25);
    \draw[Goldenrod,thick] (-3,-4) rectangle ++(5,-3);
    \draw[Goldenrod,thick] (-2.25,-4.2) rectangle ++(2.5,-2);
    \draw[black,thick,fill=black] (1,-4) rectangle ++(0.5,-1);
    \draw[Goldenrod,thick] (3,-4.5) rectangle ++(0.2,-0.2) node[anchor=west]{Copper shield};
    \draw[Violet,thick] (3,-5.25) rectangle ++(0.2,-0.2) node[anchor=west]{Cryoperm shield};
    \draw[black,thick,fill=black] (3,-6) rectangle ++(0.2,-0.2) node[anchor=west]{IR absorber};
\end{scope}
\end{tikzpicture}
    \caption{Circuit diagram of the measurement setup. The room temperature $R_\mathrm{b}=\SI{2000}{\ohm}$ and \SI{4}{\kelvin} cold $R_\mathrm{sh}=\SI{20}{\milli\ohm}$ forms a $10^5:1$ voltage divider that significantly reduces thermal and electromagnetic interference (EMI) noise from the lab environment. The connection from \SI{300}{\kelvin} to \SI{4}{\kelvin} uses twisted constantan wires, and the connection from \SI{4}{\kelvin} to JJ uses twisted superconducting niobium-titanium wires. A capacitor $C_\mathrm{f}=\SI{10}{\micro\farad}$ is placed at the voltage supply output to form a low-pass filter with $R_b$. Additional \SI{1}{\mega\hertz}-cutoff low-pass L-C filter is integrated in the D-sub connectors when the wires entering the DR. The readout SQUID is a Quantum Design SQUID array, simplified as a single loop in the schematic.}
    \label{fig:I-V setup}
\end{figure}

The device was mounted in a Bluefors LD-400 dilution refrigerator with a base temperature of 13~mK. A simplified wiring and filtering schematic is shown in Fig.~\ref{fig:I-V setup}. The bias current was supplied by a Keysight B2962B source operated with its low-noise filter. To suppress high-frequency noise on the bias line, a bias resistor of $R_b = 2000~\Omega$ and a shunt capacitor of $C_r = 10~\mu\mathrm{F}$ was installed to implement a low-pass filter. A shunt resistor, $R_{\mathrm{sh}} = 20~\mathrm{m}\Omega$, was placed in parallel with the junction to establish a voltage bias across the device.

We used NbTi twisted wires for the connection between $R_{\mathrm{sh}}$ and the junction to minimize parasitic series resistance. The junction current was read out with a single-channel DC SQUID system from Quantum Design~\cite{QDSQUID}. The SQUID input coil had an inductance of $L_{\mathrm{in}} = 2~\mu\mathrm{H}$, and the input-referred current noise was approximately $1~\mathrm{pA}/\sqrt{\mathrm{Hz}}$. Output of the SQUID amplifier was recorded with NI-9239 data acquisition card.

\section{T$_C$ Measurements}\label{Sup:Tc}
To measure the superconducting transition temperature ($T_c$), we used a Quantum Design PPMS DynaCool system equipped with an adiabatic demagnetization refrigerator (ADR). Hf films were first deposited with titanium/gold (Ti/Au) contacts by e-beam evaporation using a shadow mask to define the contact geometry for four-wire resistance measurements. The samples were then wire-bonded to a measurement puck and mounted in the ADR system. The system was cooled down to 90~mK, and the resistance was measured as a function of temperature during the warmup cycle. The puck is in thermal contact with the ADR system, which provides accurate temperature readout. We employed the standard four-point probe method with a measurement current of 1~\textmu A. The $T_c$ was determined as the temperature corresponding to 50\% of the normal-to-superconducting transition step. Table~\ref{tab:tab_S1} shows the measured and extracted parameters of the 30~nm and 60~nm thick Hf films.

\begin{table}[h!]
\centering
\caption{\label{tab:tab_S1}Summary of measured parameters for Hf thin films.}
\begin{tabular}{lcc}
\hline
\textbf{Film Thickness} & \textbf{30 nm} & \textbf{60 nm} \\
\hline
$T_c$ (mK) & 298 & 268 \\
$\Delta$ ($\mu$eV) & 41 & 44 \\
Resistivity ($\mu\Omega\,$cm) & $73.4$ & $62.1$ \\
RRR & 1.68 & 1.97 \\
\end{tabular}
\label{tab:Hf_parameters}
\end{table}

\end{document}